\documentclass[
	a4paper, % Paper size, use either a4paper or letterpaper
	10pt, % Default font size, can also use 11pt or 12pt, although this is not recommended
	twoside, % Two side traditional mode where headers and footers change between odd and even pages, comment this option to make them fixed
]{LTJournalArticle}

\usepackage{algpseudocode}
\usepackage{algorithm}
\usepackage{booktabs}
\usepackage{tabularx}
\usepackage{amsmath}
\usepackage{hyperref}
\usepackage[counterclockwise, figuresleft]{rotating}
\usepackage{placeins}

\runninghead{Multi-Modal Community Detection} % A shortened article title to appear in the running head, leave this command empty for no running head

\title{Community detection in multi-modal data:\\ A similarity network perspective} % Article title, use manual lines breaks (\\) to beautify the layout

\author{%
	Aidan Marnane\textsuperscript{1}\thanks{Corresponding author: \href{mailto:aidan.marnane@gmail.com}{aidan.marnane@gmail.com}\\ \textbf{Published:} February, 2025} \, and T. Ian Simpson\textsuperscript{1} 
}

\date{\footnotesize\textsuperscript{\textbf{1}}School of Informatics, University of Edinburgh, 10 Crichton Street, EH8 9AB, Edinburgh, UK}

\renewcommand{\maketitlehookd}{%
	\begin{abstract}
		\noindent Similarity network construction is a fundamental step in many approaches to community detection in biomedical analysis. It is utilised both in the creation of network structures from non-relational data and as a processing step in clustering pipelines. The foundation of any network analysis approach hinges on the quality of the underlying network. With the rising popularity of network learning and use of network-based clustering, the importance of correctly constructing the network is vital. The underlying mechanisms of similarity network construction, particularly the implications of the choice of approach for multi-modal integration, remain poorly explored. By introducing differences in embedded cluster information and noise levels across modalities, we assess the performance of popular similarity integration techniques such as Similarity Network Fusion (SNF) and NEighborhood based Multi-Omics clustering (NEMO). Notably, SNF and NEMO fail to outperform simpler techniques such as mean similarity aggregation when incorporating modalities with inconsistently embedded clusters. We demonstrate how integration methods can be used to incorporate partial modalities - datasets where not all individuals have a full set of measurements in all modalities. SNF shows significant sensitivity to incomplete modalities while NEMO and mean aggregation are more resilient. %\\
	\end{abstract}
}

\begin{document}

\maketitle % Output the title section

%----------------------------------------------------------------------------------------
%	ARTICLE CONTENTS
%----------------------------------------------------------------------------------------

\section{Introduction}
\label{3sec:intro}
In the digital age, we are witnessing an influx of diverse data in various forms, presenting both exciting opportunities and significant challenges \cite{mirzaMachineLearningIntegrative2019, jordanMachineLearningTrends2015}. Effectively incorporating and utilising this wealth of data is a complex task, given the different properties and challenges associated with various data types. 

Biomedical data poses distinctive challenges due to its multi-modal nature, encompassing various forms ranging from high-dimensional multi-omic data capturing genetic information facets like RNA gene expression, DNA methylation sites, and copy number variants, to diverse medical data types, including images and clinical information derived from diagnostic questionnaires \cite{santiago-rodriguezMultiOmicData2021, acostaMultimodalBiomedicalAI2022}. This data exhibits a characteristic combination of a small number of observations and high dimensionality. In addressing this unique landscape, one particularly successful approach is the utilisation of similarity networks for multi-view learning.

By extracting the relationships within datasets using similarity measures, similarity network approaches are effective in overcoming the challenges posed by the high ratio of features to observations, allowing for a nuanced exploration of specific biomedical applications. What sets similarity networks apart is their high interpretability and adaptability, serving as versatile data structures suitable for both unsupervised and supervised tasks. These tasks range from community detection to node/edge prediction, making similarity networks a valuable tool for uncovering insights in biomedical data \cite{fortunatoCommunityDetectionNetworks2016, suNetworkEmbeddingBiomedical2018}.

One of the most successful and widely adopted techniques for constructing multi-modal similarity networks is Similarity Network Fusion (SNF) \cite{wangSimilarityNetworkFusion2014}. Despite its success, the original assessment metrics used to evaluate SNF's performance were not conventional clustering accuracy measures such as Adjusted Rand Index (ARI) or Adjusted Mutual Information. Instead, indirect metrics, such as differences in survival rates between clusters and the number of significant genes within clusters, were employed. A primary challenge in evaluating SNF was the absence of data with known ground truth clusters. Determining the optimal conditions for SNF, or when simpler methods like mean similarity are sufficient, remained unclear. Notably, a study employing formal measures such as ARI found that mean similarity consistently outperformed SNF \cite{mitraMultiviewClusteringMultiomics2020}. 

To address this ambiguity, we introduce a framework for generating multi-modal data with straightforward variations in distribution and embedded cluster information. These variations enable us to assess how differences in the consistency of individual similarities across modalities impact the community detection performance of integration methods. This framework offers a systematic approach to evaluating the effectiveness of various integration methods under controlled conditions, shedding light on the circumstances where simpler methods may suffice or where the complexity of SNF proves advantageous.

A notable characteristic of biomedical data is the presence of partially complete modalities. In multi-modal datasets, it is rare to have a complete set of measurements for all individuals. \textit{Unit non-response}, where individuals have no measured features, is a frequent occurrence. In uni-modal analyses, these individuals are typically excluded from the study. However, in multi-modal data, this practice can result in significant data wastage, leading to the exclusion of large numbers of individuals or entire modalities to maximise observations. Similarity networks are well suited to mitigating data wastage by incorporating partial data. \\

In summary, in this paper we demonstrate the effect of similarity integration approaches on community detection performance of constructed networks. A challenge in the evaluation of network construction is the lack of datasets with known ground truth community structure and data properties in each modality. To overcome this we use multiple instances of synthetic data to allow an exploration of different levels of noise and consistency across modalities. We evaluate several network community detection methods across a range of different cluster settings. Section \ref{3sec:background} describes the similarity integration methods used in the construction of similarity networks. Section \ref{3sec:multimodgen} describes our multi-modal data generation framework, how we embed different cluster information, the different data distributions used and our approach to creating partial data. In Section \ref{3sec:exp} we introduce the experiments before showing and discussing the results produced in Sections \ref{3sec:results} and \ref{3sec:discussion}.

%------------------------------------------------

% \section{Methods}

\section{Multi-Modal Similarity}\label{3sec:background}
\begin{figure*}[!htbp] % Single column figure
	\includegraphics[width=0.82\linewidth]{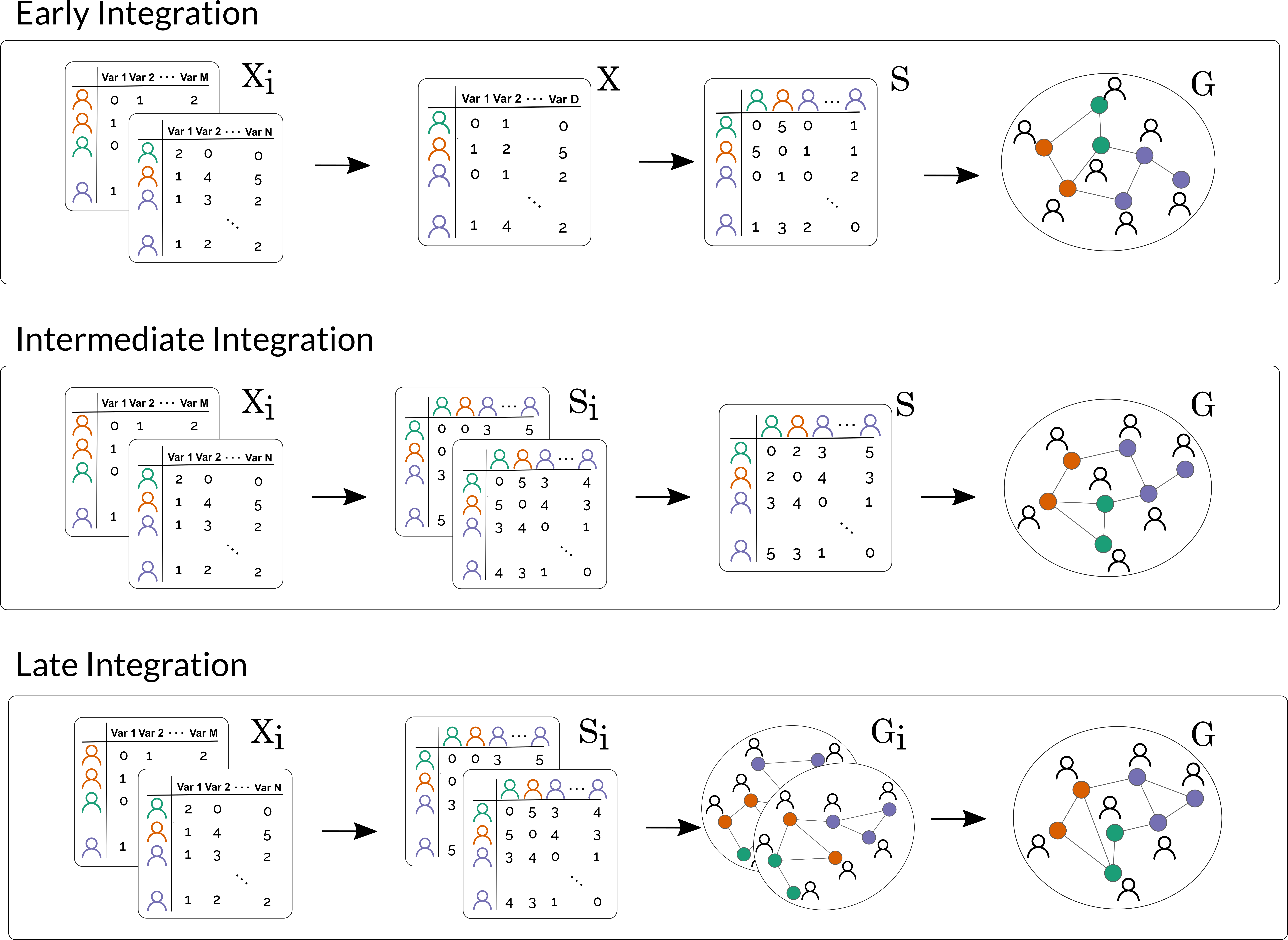}
    \centering
	\caption{\textbf{Approaches to Similarity Integration in Multi-Modal Network Construction.} Methods can be classified as early, intermediate or late integration techniques where one of the modality's i) data features $X_i$, ii) pairwise similarities $S_i$ or iii) individual networks $G_i$ are integrated together in order to construct a similarity network $G$ for the dataset.}
	\label{fig:netintegration}
\end{figure*}

In this paper, we evaluate the quality of the network produced from multi-modal data by similarity integration methods, specifically when applied to community detection. As shown in Figure \ref{fig:netintegration} integration methods can be categorised as early, intermediate and late. Our aim is to evaluate and identify which integration methods are most beneficial in the specific context of community detection. We consider five integration approaches
% \begin{itemize}
%     \item \textbf{Concatenated $X_i$} --- All modalities are combined into a single feature matrix $X$. Pairwise similarity and network sparsification are subsequently performed.
%     \item \textbf{Mean $S_i$} --- Mean similarity between a pair of nodes $i$ and $j$ across all modalities.
%     \item \textbf{Extreme Mean} --- Mean "extreme" similarity/dissimilarity between a pair of nodes $i$ and $j$ across all modalities. For each modality, pairwise similarity is thresholded to only include very similar and very dissimilar connections.
%     \item \textbf{Similarity Network Fusion (SNF)} --- \textit{de facto} standard approach for multi-omic integration and unsupervised clustering analysis. Similarity calculated through diffusion across KNN graphs.
%     \item \textbf{Neighbourhood Based Multi-Omic Clustering (NEMO)} --- Mean relative similarity between nodes $i$ and $j$ based on a K-nearest Neighbourhood in each modality.
% \end{itemize}
\begin{itemize}
    \item \textbf{Early}
    \begin{itemize}
    \item \textbf{Concatenated $X_i$} --- All modalities are combined into a single feature matrix. Afterward, pairwise similarity and network sparsification are performed.
    \end{itemize}
    \item \textbf{Intermediate}
    \begin{itemize}
    \item \textbf{Mean $S_i$} --- Mean similarity between a pair of nodes $i$ and $j$ across all modalities.
    \item \textbf{Extreme Mean} --- Mean "extreme" similarity/dissimilarity between a pair of nodes $i$ and $j$ across all modalities. For each modality, pairwise similarity is thresholded to include only very similar and very dissimilar connections.
    \end{itemize}
    
    \item \textbf{Late}
    \begin{itemize}
    \item \textbf{Similarity Network Fusion (SNF)} --- \textit{de facto} standard approach for multi-omic integration and unsupervised clustering analysis. Similarity calculated through diffusion across KNN graphs.
    \item \textbf{NEighborhood Based Multi-Omic Clustering (NEMO)} --- Mean relative similarity between nodes $i$ and $j$ based on a K-nearest Neighbourhood in each modality.
    \end{itemize}
\end{itemize}

\subsection{Concatenating Features}
The simplest approach to integrating multi-modal data is to concatenate the features from each modality into a single "master" feature matrix. This is the prototypical example of early integration. From a set of $m$ modality feature matrices $X_i$, a single data feature matrix $X$ is constructed as follows:
\begin{equation}
X = [X_1, X_2, \ldots, X_m]    
\end{equation}
The benefit of this approach lies in its simplicity and the unadjusted inclusion of each modality. If the features of any particular modality are informative, this should be captured through the similarity calculation. In practice, modalities can have highly different scales of dimensionality --- from clinical data with tens of features to DNA methylation data with hundreds of thousands of features \cite{tomczakReviewCancerGenome2015}. In such cases, the higher-dimensional modality will dominate the differences between individuals.

\subsection{Mean Similarity}
Perhaps the simplest method of multi-modal similarity integration beyond feature concatenation is to calculate similarity for each modality independently and integrateilarity scores before constructing a final network $G$. The pairwise similarity matrix\footnote{This notation is used to keep consistency with the original notation used in deriving the SNF method. With respect to Figure \ref{fig:netintegration}, the corresponding notation is $P^{(v)} = S_v$} $P$ is given by
\begin{equation}
    P_\text{Mean} = \frac{1}{m} \sum_{v=1}^{m}P^{(v)}
\end{equation}
where $m$ is the number of modalities and $P^{(v)}$ is the pairwise similarity for modality $v$. This intermediate integration technique involves calculating the pairwise distances independently for each modality and then merging them to form a single pairwise similarity matrix. From this final pairwise similarity matrix, a network can be created. The benefit of this approach is the ability to process each modality independently. For example, similarity within each modality can be calculated using separate similarity measures. Each measurement of similarity is equally valued, ensuring that modalities with lower dimensionality will not be obscured by modalities with a high number of features. 

\subsection{Extreme Mean}
Typically, the focus in community detection is on extreme similarities or dissimilarities, as relationships between nodes that are mildly similar or dissimilar are often considered uninformative in network construction. These less informative relations are usually filtered out during the sparsification process. Connections between nodes that exhibit high similarity form communities within the network. However, in realistic scenarios such as disease analysis, negative relations (high dissimilarity) can be crucial. The distinct dissimilarity between two individuals in a subset of features or measurements can provide strong evidence that these individuals are not alike and likely do not share the same disease or disease subtype. Connecting these dissimilar individuals in network construction would be inaccurate. One approach to creating a network based on strong (dis)similarities is to threshold the similarities of each modality before integration. 

To threshold a modality's pairwise similarity matrix $P$, we apply the following rule:
\begin{equation}
   W(i,j) = \begin{cases}
P(i,j), & \text{if} \; |P(i,j)| > \sigma\\
0, & \text{if} \; |P(i,j)| < \sigma
\end{cases}
\end{equation}
where $\sigma$ is chosen threshold value. This thresholding process can be straightforwardly applied to normalised similarity metrics, such as Pearson correlation. After thresholding, only highly positively correlated or negatively correlated relationships will be retained. For unnormalised metrics like Euclidean distance, one can normalise the pairwise distances to obtain a zero mean, identity standard deviation distribution of pairwise distances, making it easier to choose an interpretable threshold value. In this work, a threshold of $\sigma=1$ standard deviation is used, ensuring that only pairwise values that are significantly similar or dissimilar are retained.

To obtain the final pairwise similarity matrix ($P_{\text{Extr}}$), we compute the mean of the per-modality thresholded pairwise similarities:
\begin{equation}
P_{\text{Extr}} = \frac{1}{m} \sum_{v=1}^{m}W^{(v)}    
\end{equation}
This method can be considered as both an intermediate and late integration method. The threshold step is akin to threshold sparsification\footnote{Also referred to as $\epsilon$-networks.} where edges are included if their similarity is above a particular threshold, making it a form of integration of weighted threshold networks. Unlike threshold sparsification, this method retains highly dissimilar connections, and the resulting weighted pairwise similarities after thresholding do not represent a typical threshold network.

\subsection{Similarity Network Fusion}
Similarity Network Fusion (SNF) \cite{wangSimilarityNetworkFusion2014} is a late integration approach for constructing a multi-modal similarity network. SNF employs an iterative diffusion process to converge on a single pairwise similarity matrix. The primary goal of SNF is to update the similarity between two nodes (in any given modality) based on the similarity of their shared nearest neighbours across all modalities. It can be thought of as creating a weighted K-nearest neighbour (KNN) network for each modality and merging these networks by adjusting the weights between nodes based on their shared neighbours. Therefore, SNF is a late integration method that combines networks in a non-linear manner.

In \cite{wangSimilarityNetworkFusion2014}, Wang \textit{et al.} introduce the two key components of SNF: i) a scaled exponential similarity kernel to compute affinity (similarity) between all nodes on each modality and ii) a diffusion process to merge the similarity for separate modalities. The diffusion step was a key contribution for modality integration and the method was validated through the identification of cancer subtypes on multi-omic cancer data from The Cancer Genome Atlas (TCGA). 

% A pair of nodes similarity is given by the sum of   to share information between the nearest neighbours in  
The pairwise scaled exponential similarity kernel computed between all nodes is given by
\begin{equation}\label{3eq:snfaffinity}
    W(i,j) = \text{exp}\left( - \frac{d^2(x_i,x_j)}{\mu \epsilon_{i,j}}\right)
\end{equation}
where $d(x_i,x_j)$ is a distance metric (in the original paper, the euclidean distance was used), $\mu$ is a hyperparameter that controls which distances can be considered highly similar (for a fixed distance between nodes $i$ and $j$, lowering $mu$ lowers their similarity), and $\epsilon_{i,j}$ is a scaling factor that incorporates the distance between the nearest neighbours of $i$ and $j$. The scaling factor $\epsilon_{i,j}$ is given by
\begin{equation}
    \epsilon_{i,j} = \frac{\text{mean}(d(x_i, N_i) + \text{mean}(d(x_j, N_j) + d(x_i,x_j)}{3}
\end{equation}
where $N_i$ is the set of neighbours of node $i$. This scaled similarity kernel is qualitatively different to the euclidean distance. It is normalised between 0 and 1 and more importantly, computed values between nodes are typically either quite close to 1 or quite close to 0. Moreover, the scaled affinity kernel controls for different areas of density in the feature space. For a node $x_i$, if $d(x_i, N_i)$ is large on average then $d(x_i, x_j)$ being large will not be as penalised. In other words, if a cluster is more spread apart, the pairwise similarity of its nodes will be as high as a more tightly knit cluster. 

There are a number of key normalisations that are performed before the diffusion process to ensure numerical stability. A normalised weighed pairwise affinity for each modality $P$ is created by
\begin{equation}
   P(i,j) = \begin{cases}
\frac{W(i,j)}{2 \sum_{k\neq i} W(i, k)}, & j \neq i\\
1/2, & j = i
\end{cases}
\end{equation}
which ensures that $\sum_j P(i,j)=1$. A normalised weighted KNN network with adjacency matrix $S$ for each modality
\begin{equation}
   S(i,j) = \begin{cases}
\frac{W(i,j)}{\sum_{k \in N_i} W(i, k)},& j \in N_i\\
0 & \text{otherwise}
\end{cases}
\end{equation}
where $N_i$ is the set of neighbours of node $i$.

To integrate the modality together diffusion is performed across the KNN networks using the similarity in other modalities. The diffusion step for modality $v$ is given by
\begin{equation}
    P^{(v)} = S^{(v)} \times \left( \frac{\sum_{u \neq v} P^{(u)}}{m - 1} \right) \times (S^{(v)})^T, v = 1,2,\dots,m   
\end{equation}
The similarity $P^{(v)}$ for each modality $v$ is updated by considering the similarity of the local neighbourhood of nodes $i$ and $j$ in other modalities (the only non-zero elements in $S^{(v)}$ are the nearest neighbours of a node). 

This can be more clearly seen if we express the update rule from iteration $t$ to $t+1$ as
\begin{align}
    P_{t+1}^{(v)}(i,j) = \sum_{k \in N_{v_i}} \sum_{l \in N_{v_j}} S^{(v)}(i,k) & \times S^{(v)}(j,l) \nonumber  \\ 
    & \times \left( \frac{\sum_{u \neq v} P_{t}^{(u)}(k,l)}{m - 1} \right)
\end{align}
If the neighbours of node $i$ and $j$ are highly similar in other modalities ($P_t^{(u)}(k,l)$), then $P^{(v)}(i,j)$ will increase and the edge $(i,j)$ is more likely to be included in the final network. Conversely, if the neighbours of node $i$ and $j$ are very unsimilar, this reduces the evidence of a relationship between nodes $i$ and $j$. $P^{(v)}(i,j)$ will decrease, making it less likely the edge $(i,j)$ will be included in the network. 

After each iteration the modalities are re-normalised (for numerical stability) until convergence is achieved. The final network pairwise similarity is given by 
\begin{equation}
    P_\text{SNF} = \frac{1}{m} \sum_{v=1}^{m}P^{(v)}
\end{equation}
Typically, very few iterations are required. The authors found only 1 or 2 needed to converge on the TCGA data. 

% % problem with method
% SNF was developed on and applied to multi-omic cancer sets with no known ground truth clusters. As a result, the accuracy of the method had to be evaluated indirectly. Differences in the survival rate of the subtypes and the number of significant genes present in each subtype were used as evidence of the success of the method. It was not evaluated with typical clustering metrics such as ARI or AMI. There still remains unanswered questions over what types data the method is best suited to? Does SNF always outperform simpler approaches such as Mean $S_i$?

% SNF was developed on and applied to multi-omic cancer datasets without known ground truth clusters. As a result, the accuracy of the method had to be evaluated indirectly. Differences in the survival rate of the subtypes and the number of significant genes present in each subtype were used as evidence of the success of the method. However, it was not evaluated with typical clustering metrics such as Adjusted Rand Index (ARI) or Adjusted Mutual Information (AMI). This leaves unanswered questions regarding the types of data the method is best suited for. Does SNF consistently outperform simpler approaches such as Mean $S_i$?

\subsection{NEMO}
NEMO (Neighbourhood based Multi-Omics clustering) is an alternative approach to similarity integration \cite{rappoportNEMOCancerSubtyping2019}. Similar to SNF, it is a late integration approach that combines networks from each modality. It is a simpler approach than SNF. NEMO does not make use of diffusion and instead integrates a nodes neighbourhood information by creating a KNN network on each modality. A final network is created by computing a weighted sum of the individual networks. Similar to SNF, NEMO is a late integration method that combines networks rather than pairwise distances in order to create a final network. 

% S^{(v)}
Initially for each modality, a KNN network is created from their pairwise similarity matrix and the relative similarity between nodes is calculated using 
\begin{align}
    S^{(v)}(i,j) =& \frac{W^{(v)}(i,j)}{\sum_{r \in N_{v_i}} W^{(v)}(i,r)} \cdot I(j \in N_{v_i}) \nonumber \\
    & + \frac{W^{(v)}(i,j)}{\sum_{r \in N_{v_j}} W^{(v)}(r,j)}\cdot I(i \in N_{v_j})
\end{align}
where $N_{v_i}$ is the set of neighbours of node $i$ in modality $v$, $W^{(v)}$ is the pairwise affinity between nodes. Similar to SNF, NEMO makes use of the scaled exponential affinity kernel (Eq. \ref{3eq:snfaffinity}).

Finally, the average similarity is calculated using
\begin{equation}
    P_\text{NEMO} = \frac{1}{m} \sum_{v=1}^{m}S^{(v)}
\end{equation}

For partial data the average relative similarity is adjusted to only take the mean of the modalities where both nodes are present
\begin{equation}
P_\text{NEMO} = \frac{1}{|\sigma_{ij}|} \sum_{v \in \sigma_{ij}}S^{(v)}(i,j)    
\end{equation}
where $\sigma_{ij}$ is the set of modalities where $i$ and $j$ are both present. In other words, the similarity between two nodes is calculated ignoring missing values and does not "punish" the similarity between two nodes if they have modalities where they are not present. Unlike other approaches, the increased uncertainty in similarity between nodes with missing modalities does result in a reduced level of similarity.

% why good alternative

% \begin{equation}
%     RS_l(i,j) = \frac{S_l(i,j)}{\sum_{r \in N_{l_i}} S_l(i,r)} \cdot I(j \in N_{l_i}) + \frac{S_l(i,j)}{\sum_{r \in N_{l_j}} S_l(r,j)}\cdot I(i \in N_{l_j})
% \end{equation}

% \begin{equation}
%     ARS = \frac{1}{L} \sum_l RS_l
% \end{equation}

% talk about data? 

% \FloatBarrier
% %%%%%%%%%%%%%%%%%%%%%%%%%%%%%%%%%%%%%%%%%%%%%%%%%%%%%%%%%%%%%%%%%%%%%%%%%%%%%%%%%
\FloatBarrier
\section{Synthetic Data}\label{3sec:multimodgen}
A key assumption in multi-modal data analysis is that each modality captures a different aspect or view of the underlying community structure. Another assumption is that different modalities have different distributions and properties that require individual processing. With these factors in mind, a "good" data generator will allow the adjustment of both the underlying data distribution and the embedded cluster information of each modality.

In this work, we propose a framework for the generation of high-dimensional data where the distribution and cluster information in each modality can be adjusted separately. Each individual modality's clustering problem is non-trivial and the performance of different multi-modal similarity integration techniques can explored in detail. The data generation method proposed here scales to a high number of modalities. 

Our generation procedure is as follows i) generate ground truth set of clusters $y$, ii) for each modality $i$, we generate the modality cluster ground truth $y_i$ derived from $y$ and iii) for each modality $i$, we generate the data $X_i$ from the modality clusters $y_i$. We split the population of nodes $N$ into clusters of equal size.  While less realistic than varying cluster sizes, equally sized clusters allow improved evaluation by isolating the effect of changes in modality distribution and cluster information. It should be noted that this generation procedure facilitates clusters of any size. 

\begin{figure*}[!htbp] % Single column figure
	\includegraphics[width=0.75\linewidth]{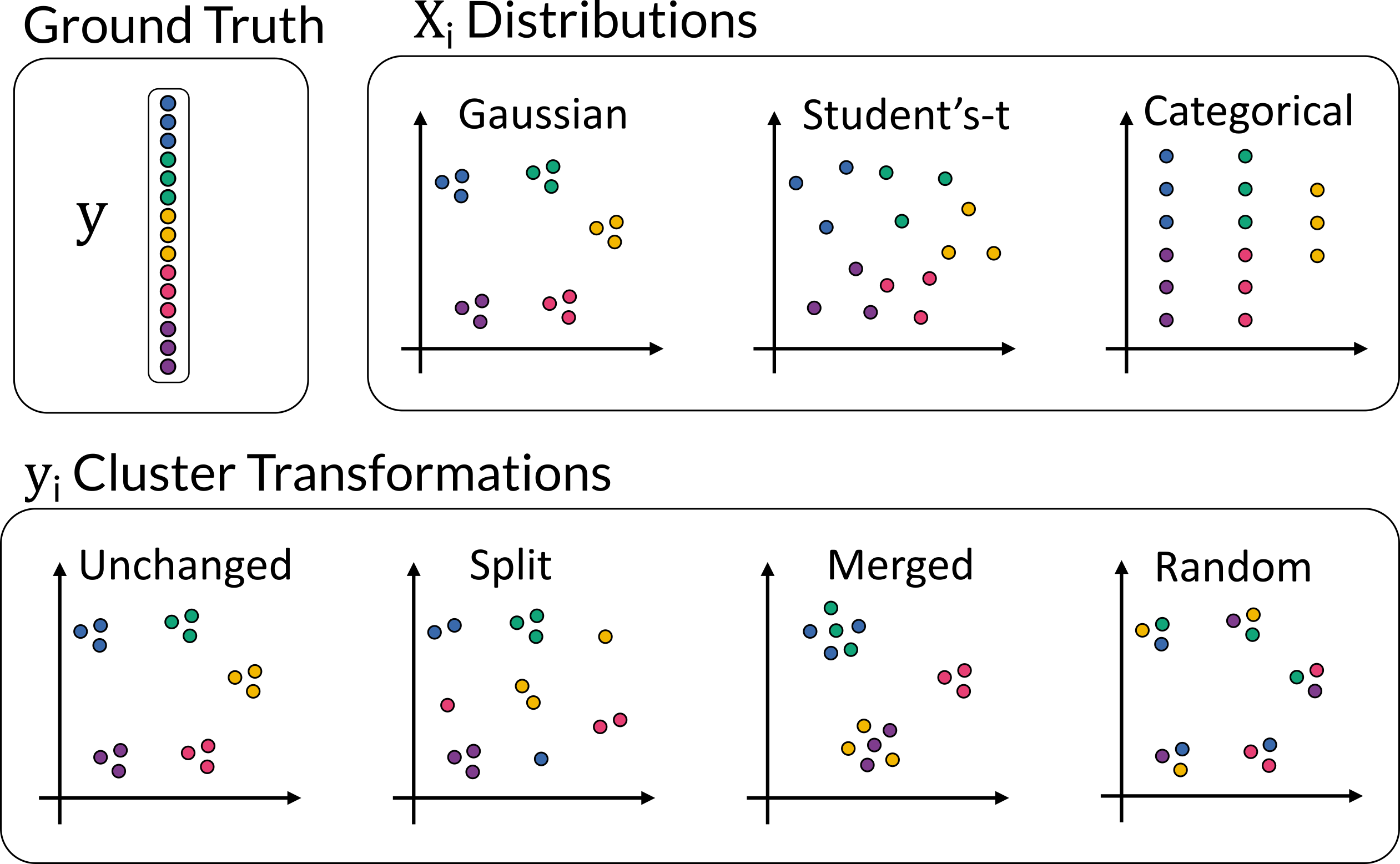}
    \centering
	\caption{\textbf{Generation of Modality-Specific Clusters and Feature Distributions.} This figure illustrates the possible components that can be adjusted in the process of generating modality-specific clusters and features from the ground truth labels $y$. For each modality $i$, the modality ground truth clusters $y_i$ are derived by applying one of four transformations to $y$: (i) keeping $y_i$ identical to $y$, (ii) splitting clusters in $y$ into subclusters, (iii) merging clusters in $y$, or (iv) generating random, unrelated clusters. Features $X_i$ are then generated based on $y_i$ using one of three distributions: (i) mixture of Gaussians, (ii) mixture of Student's-t, or (iii) categorical data. }
	\label{3fig:modality_types}
\end{figure*}

As shown in Figure \ref{3fig:modality_types}, we can adjust both the distribution of the data $X_i$ and the method of generating the modality clusters $y_i$ from $y$. We want each modality to capture a different aspect of the community structure. To replicate this in our dataset we require a method of adjusting the embedded clusters in each modality while still ensuring a ground true community structure across the entire dataset. Our proposed solution is to sample a set of ground truth cluster labels $y$ for all modalities and generate per modality labels $y_i$ from $y$ that control how individuals are distributed across embedded clusters within a particular modality. We want these embedded clusters to differ from the ground truth while remaining consistent, for example, we embed 3 clusters in $X_0$ but ensure each cluster in $y_0$ is assembled from the 5 clusters in $y$. We also adjust the data distribution of the features ($X_i$) of the clusters in each modality ($y_i$) by selecting from one of three distributions. 

\subsection{Distributions}
We propose using three types of cluster distributions as shown in Figure \ref{3fig:modality_types} 
\begin{itemize}
    \item Mixture of Gaussians,
    \item Mixture of Student's-t,
    \item Categorical Data.
\end{itemize}
The Mixture of Gaussians is a commonly assessed distribution with implanted community structure. Members of each individual cluster are sampled from separate high dimensional Gaussian distributions. We assign each cluster identity covariance so that the sole difference between clusters are the locations of the center of each distribution. Cluster centers are generated in such as way as to ensure some level of overlap between clusters at higher dimensions. This overlap prevents trivial detection of clusters and ensures a challenging cluster problem in each modality. The Mixture of Student's-ts is a noisier, more challenging variant of the Mixture of Gaussian distributed data. We sample from high dimensional Student's-t distributions with 2 degrees of freedom, identity covariance and unique centers for each cluster. Outliers are far more likely and the level of overlap between clusters is increased.

The categorical data distribution is comprised of a mix of informative and uninformative features. In an informative feature, each cluster has an individual probability distribution across $n$ possible categories and the value of each member of the cluster is sampled according to that distribution. In an uninformative feature, each individual in the dataset samples according to a shared distribution across the $n$ categories. This distribution is highly consistent from modality to modality. The discrete number of possible values each cluster can take ensures that outliers are highly unlikely and within clusters distances do not vary significantly from one modality to the next.

\subsection{Cluster Information}
 We want to generate data that allows us to evaluate scenarios; i) where the cluster distributions changes from modality to modality and ii) where the cluster information is consistent. To adjust the cluster information embedded in each modality we propose three methods of adjustment.
\begin{itemize}
    \item Splitting $y$ into sub-clusters
    \item Merging clusters in $y$ into super-clusters
    \item Generating random set of clusters unrelated to $y$
\end{itemize}

In Figure \ref{3fig:modality_types}, we illustrate these methods of adjusting $y_i$ on a simple two dimensional example. Our aim in making these adjustments is to evaluate the ability of similarity integration methods to handle inconsistencies across modalities. Splitting and merging $y$ into sub and super clusters is most reflective of real world settings were possible subtypes or cell types likely have differences in some modalities but share traits in others. While the cluster distribution is not identical to the true cluster distribution it will still be quite consistent from one modality to the next on less noisy distributions. \\

By including a random modality where the cluster distribution $y_i$ is unrelated to $y$ we are able to assess the ability of integration methods to  not just handle noise within a dataset but also handle the inclusion of uninformative data. There is no guarantee in real world settings that all modalities will be informative for example a particular disease may affect an individuals transcriptomic data but not its genomic data and so the inclusion of genomic data only adds noise to the dataset. In particular, the random modality add significant inconsistencies for the middle/late integration methods as the inter cluster distances and KNN networks generated from random modalities are guaranteed to be unlike the ground truth data. \\

\subsection{Generating Partial Data}
A number of studies have analysed the problem of partially complete data and its affect on the accuracy of multi-modal methods \cite{rappoportNEMOCancerSubtyping2019, liPartialMultiViewClustering2014, xuDeepIncompleteMultiView2022}. However, these methods have only analysed the possibility of data being partial at random and have not analysed the effect of partial data on network structure. We define partially complete data to refer to a multi-modal dataset with $m$ modalities where a subset of the individuals have no recorded data in at least 1 of the $m$ modalities and a complete set of measurements in the other modalities. Recall that this differs from a more typical missing data scenario where a much smaller proportion of individuals are missing values in a subset of the $d$ features within a dataset. \\

\begin{figure}[!htbp] % Single column figure
    \centering
	\includegraphics[width=0.95\linewidth]{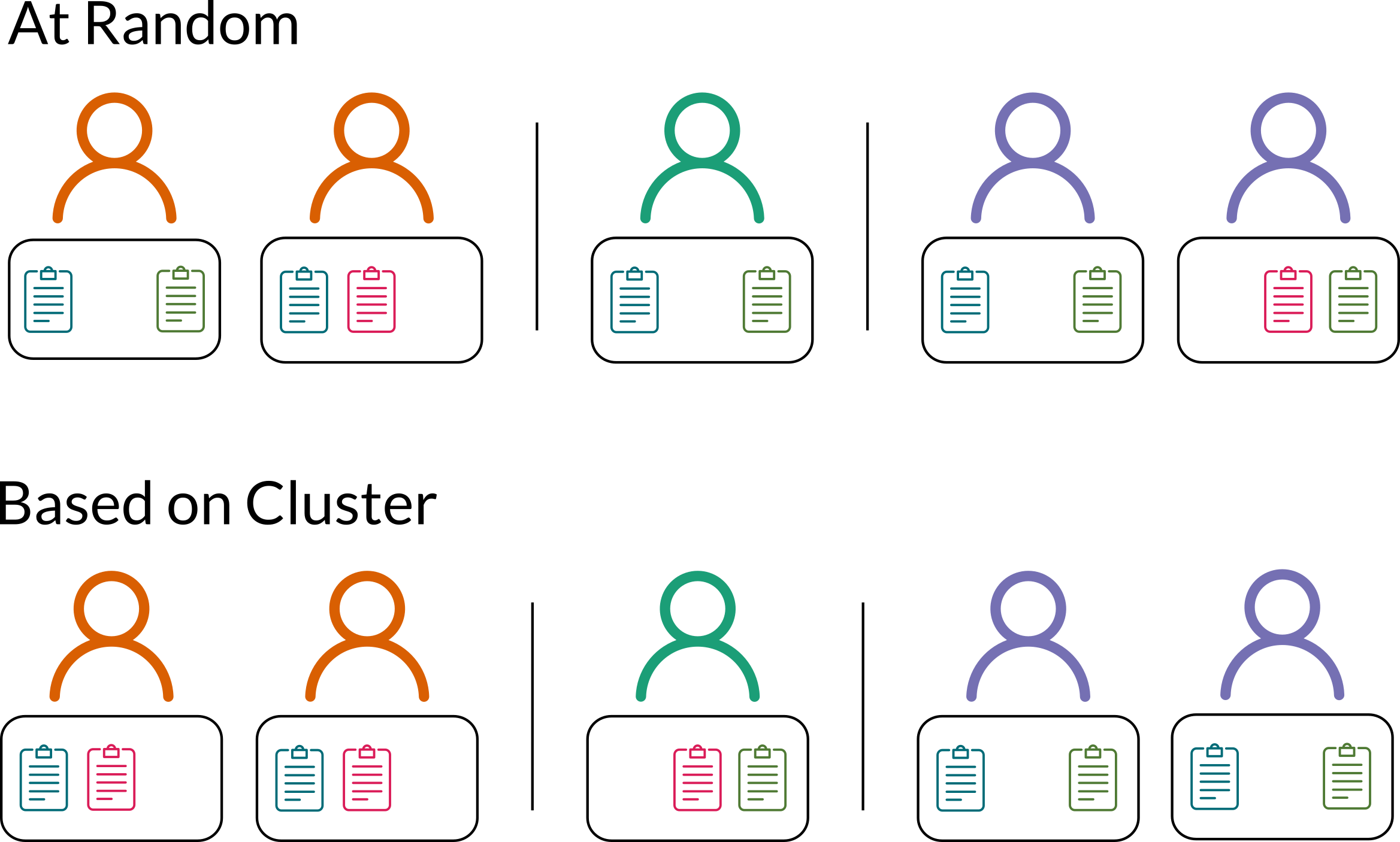}
	\caption{\textbf{Types of Partial Data in Multi-Modal Datasets} This figure illustrates two scenarios of partial data in multi-modal datasets: missing data either at random or based on cluster membership. When measurement are missing based on cluster, only individuals from cluster 1 (orange) do not have measurements in modality 3 (light green). In data partial at random, there is no link between the cluster label and the partial data.}
	\label{1fig:partial_data_reasons}
\end{figure}

As shown in Figure \ref{1fig:partial_data_reasons}, we want to assess two types of partial data scenarios; i) where the data from each modality is absent at random, i.e., each cluster is equally likely to have entities with no data recorded and ii) where the data is missing from each modality due to its cluster membership, i.e., one or more clusters have no data recorded in a modality. We restrict our partial data experiments to partial data where each individual has at most one non recorded modality. 

We generate partial datasets as follows. Firstly, a multi-modal dataset is generated (as described in Section \ref{3sec:multimodgen}. Then to create a partial dataset entities are removed from a modality either at random or based on their true cluster $y$. To create data partial at random, we randomly generate a set of labels $y_{NaN}$ where each label is $y_{NaN_i} \in \{1,\ldots,m\}$ where $m$ is the number of modalities in the dataset. Entities are removed from a modality based on their label in $y_{NaN}$. To create partial data based on cluster membership, we create a set of labels $y_{NaN}$ by merging the clusters in $y$ into $m$ super-clusters (if $m < n_{c}$ where $n_c$ is the number of clusters in $y$) or splitting $y$ into $m$ sub-clusters if ($m > n_c$). If $m=n_c$ then we set $y_{NaN} = y$. Again entities are removed from a modality based on their label in $y_{NaN}$. \\

\FloatBarrier
\section{Experiment Setup}\label{3sec:exp}
We conduct three experiments. We evaluate i) the performance of our multi-modal integration methods on a variety of different modality problems, ii) the adaptability of the integration methods to an increasing number of modalities, and iii) the ability of each integration method to incorporate partial modalities where a subset of individuals do not have features in some of the modalities. \\

\subsection{Integration Methods}
As described in Section \ref{3sec:background}, we evaluate several multi-modal similarity integration approaches.

\begin{itemize}
    \item \textbf{Similarity Network Fusion (SNF)} --- \textit{de facto} standard approach for multi-omic integration and unsupervised clustering analysis. Similarity calculated through diffusion across KNN graphs.
    \item \textbf{Neighbourhood Based Multi-Omic Clustering (NEMO)} --- Mean relative similarity between nodes $i$ and $j$ based on a K-nearest Neighbourhood in each modality.
    \item \textbf{Mean $S_i$} --- Mean similarity between a pair of nodes $i$ and $j$ across all modalities.
    \item \textbf{Extreme Mean} --- Mean "extreme" similarity/dissimilarity between a pair of nodes $i$ and $j$ across all modalities. for each modality, pairwise similarity is thresholded to only include very similar and very dissimilar connections.
    \item \textbf{Concatenated $X_i$} --- All modalities are combined into a single feature matrix. Pairwise similarity and network sparsification are subsequently performed.
\end{itemize}

We use make use of a python implementation of \textit{SNF}, the \texttt{snfpy}\footnote{\href{https://github.com/rmarkello/snfpy} --- v0.2.2} package. We use custom python implementations for the \textit{NEMO}, \textit{Mean $S_i$}, \textit{Extreme Mean} and \textit{Concatenated $X_i$} similarity integrators. The final network produced by all similarity integration methods is created by constructing a K-nearest Neighbour (KNN) Graph\footnote{Data with 2500 individuals is evaluated. $K=25$ was found to produce networks of a desirable density.} with $K=25$. For the SNF affinity function (Eq. \ref{3eq:snfaffinity}), we use the original proposed hyperparameter settings for the SNF Kernel $\mu=0.5$ and a value $K=25$ to match the final KNN graph. Unless otherwise specified in the creation of the pairwise similarity matrix, we make use of raw distance for the \textit{Concatenated $X_i$} and \textit{Mean $S_i$} methods, and the SNF affinity function for \textit{SNF}, \textit{NEMO} and \textit{Extreme Mean}. \\

\subsection{Clustering Algorithms}
We perform community detection on the multi-modal graph networks using three distinct network clustering algorithms.
\begin{itemize}
    \item \textbf{SBM} --- Micro-canonical Stochastic Block Model \cite{peixotoNonparametricWeightedStochastic2018}. Python \texttt{graph-tool}\footnote{v2.45} implementation \cite{peixotoGraphtoolPythonLibrary2014}.
    \item \textbf{Leiden} --- Modularity maximisation using Leiden algorithm \cite{traagLouvainLeidenGuaranteeing2019}. Python \texttt{igraph}\footnote{v0.10.3} implementation \cite{csardiIgraphSoftwarePackage2006}. The resolution hyperparameter is selected using event sampling \cite{jeubMultiresolutionConsensusClustering2018}. 
    \item \textbf{Spectral} --- Spectral decomposition and K-means clustering of "Random Walk" normalised Laplacian $L_{rw} = I - D^{-1} A$. Python \texttt{spectralclusterer}\footnote{v0.2.16} implementation \cite{wangSpeakerDiarizationLSTM2018}.
\end{itemize}

We evaluate the quality of the networks produced by the similarity integration methods using the following network statistics
\begin{itemize}
    \item \textbf{Modularity $y$} --- Network modularity compares the observed number of edges within a set of clusters to the number of edges expected under a null model (node degrees are fixed and edges are placed at random). The modularity of a graph $G$ is given by 
    \[Q = \frac{1}{2 m} \sum_{i,j} \left[ A_{ij} - \gamma \frac{k_i k_j}{2*m} \right] \delta(C_i,C_j)\]
    where $m$ is the number of edges in $G$, $A$ is the adjacency matrix of $G$, $k_i$ is the degree of node $i$ and $C_i$ is the cluster that node $i$ belongs to. We calculate the modularity of the ground truth clusters $y$.
    \item \textbf{Triad participation ratio (TPR) $y$} --- is the fraction of nodes in cluster $C$ that belong in a triad, \\
    $f(C) = \frac{|\{ u:u \in C, \{ (v,w): v,w \in C, (u,v) \in E, (u,w) \in E, (v,w) \in E\} \neq \emptyset\}|}{n_c}$\\
    where $n_c$ is the number of nodes in cluster c. We calculate the average TPR of the ground truth clusters $y$.
    \item \textbf{Assortativity} --- the Pearson correlation coefficient of degree between pairs of nodes with an edge connecting them. It measures the propensity for edges to exist between nodes of similar degrees. An explicit definition for degree assortativity can be found in Eq. 21 in \cite{newmanMixingPatternsNetworks2003}. Values range between $[-1,1]$. A positive value indicates that nodes of similar degree connect. A negative value indicates that high degree nodes are more likely to connect to low degree nodes.
    \item \textbf{Mean Path Length} --- Mean length of shortest paths between all pairs of nodes in $G$. 
    \item \textbf{Mean degree $k$} --- Mean degree $k$ of all nodes in $G$. The degree of node $i$ is the number of edges of between node $i$ and other nodes in $G$. 
    \item \textbf{Median degree $k$} --- Median degree $k$ of all nodes in $G$.
\end{itemize}

Modularity and TPR provides a measure of how well the true community structure has been embedded in $G$. Modularity assesses how tightly knit the communities in $G$ are. Communities with higher modularity have higher internal density and are characterised by more connections within the community than connections with nodes outside it. 

TPR measures the fraction of nodes within a community that form triads. This metric draws inspiration from observations on social networks that within communities, friends of a friend tend to be common. In other words, if two individuals share a common friend, they are more likely to be friends themselves. A good community can therefore be defined to be one that contains many such friends of friends. Again this is a measure of internal density --- the more triads in a community, the denser the internal connectivity. Yang \textit{et al.} \cite{yangDefiningEvaluatingNetwork2012} showed that TPR is a reliable metric for detecting communities in real networks. \\

Assortativity, mean path length, mean and median degree capture distinct facets of the global structure within a network. Mean and median degree provide a summary of the degree distribution --- whether the degree distribution is skewed, whether there is an abundance of high or low degree nodes. Assortativity provides insight into the type of nodes that tend to interconnect. This metric helps identify whether nodes with similar degrees are more likely to be linked.

Mean path length, on the other hand, offers insight into the connectedness of communities within the network. All our networks are constructed on the same data with the same KNN hyperparameter and should have similar numbers of edges. Differences in path length are due to differences in how edges connect the global structure. A low mean path length signifies high interconnectivity between communities drawing nodes in distinct communities together. Conversely, a high mean path length suggests fewer inter-community edges and longer distances between nodes, implying greater isolation between communities. These metrics collectively provide a comprehensive overview of a network's structural characteristics.

\subsection{Integration Method Performance}\label{3sec:exp3mod}
To evaluate the similarity integration methods, we desire a mixture of various modality problems. We want to see both the effect of type of distribution of the clusters as well as the effect of differences in the cluster information present in each modality. \textit{How well do the methods handle noisy data? How well do the methods handle inconsistencies in inter and intra cluster distances from modality to modality?} We compare the performance of similarity integration methods on datasets comprised of three generated modalities. This is reflective of real world applications where a large number of modalities are not typically available/collected. For example, the TCGA multi-omic datasets used in \cite{wangSimilarityNetworkFusion2014} were comprised of mRNA expression, DNA methylation  and miRNA expression data. 

Table \ref{tab:3modproblems} shows the set of fifteen problems used to evaluate the performance of the integration methods. The settings are listed by order of average ARI performance of SBM and Leiden cluster methods on each individual modality\footnote{Spectral clustering is not included in this ranking due to its high variability.}. We briefly describe some of the key modality problems: \textit{Easy} can be thought of as the default setting --- a mixture of Gaussians generated with unchanged cluster information per modality. \textit{Cat} evaluates the effect of categorical distributed data. \textit{Noisy} evaluates a high noise setting. \textit{Split} evaluates settings with ground truth clusters broken apart across modalities. \textit{Merged} evaluates the converse where ground truth clusters are combined together. \textit{1Rand} evaluates the addition of a discordant and uninformative modality. \textit{Mixed Normal} and \textit{Mixed Noisy} evaluate effect of sets of modalities with differences in cluster information in low and high noise settings respectively. \textit{Mixed Noisy 1Rand} is the most challenging setting where each modality has high noise and one of the modalities is uninformative. 

For each modality problem, we split 2500 entities into 10 equal clusters. Each generated modality contains 50 features i.e. each $X_i$ is a $N \times d$, 2500 x 50, matrix with the distribution and cluster information as described in Table \ref{tab:3modproblems}. The merged clusters are created by randomly merging $y$ into 5 clusters. The merging is done at random and can be unequal e.g 6 clusters in $y$ merged into 1 cluster in $y_i$ with the remaining 4 clusters unchanged. The split clusters are created by splitting the clusters in $y$ into 20 clusters. Similar to the merged clusters, this process is done at random and can be unequal e.g. one cluster in $y$ split into 11 subclusters with the remaining 9 unchanged. A random modality is created by generating assigning each entity to one of 10 equally sized clusters at random and generating the dataset with the random labelling $y_i$. Each modality problem is evaluated on 20 instances to better estimate the accuracy of the similarity integration methods. 

% We repeat the evaluation using two metrics i) euclidean distance and ii) correlation. We evaluate using both the raw distance and using SNF affinity (Eq \ref{3eq:snfaffinity}).

% To improve the estimate oevaluate the performance of each similarity integration method on 20 instances 

% To improve the estimate oevaluate the performance of each similarity integration method on 20 instances 

\begin{table}[!htbp]
\begin{tabularx}{\columnwidth}{Xp{.57cm}p{.57cm}p{.57cm}X}
\cmidrule(r){1-4}
Name              & $X_1$ & $X_2$ & $X_3$ &       \\
\cmidrule(r){1-4}
Cat               & C-0   & C-0   & C-0   &                   \\
Easy              & G-0   & G-0   & G-0   &                   \\
Single Merged     & G-0   & G-0   & G-1   &                   \\
Single Noisy      & G-0   & G-0   & S-0   &    Distributions                \\
\cmidrule(lr){5-5}
Split             & G-2   & G-2   & G-2   &    G - Gaussian               \\
Mixed Normal      & G-1   & G-1   & G-2   &  S - Student's-t    \\

Merged            & G-1   & G-1   & G-1   &    C-Categorical    \\
Mixed All         & C-1   & G-1   & S-2   &   \\
Noisy             & S-0   & S-0   & S-0   &   \\
1Rand             & G-0   & G-0   & G-3   &   $y_i$ clusters                 \\
\cmidrule(lr){5-5}
Mixed Noisy       & S-1   & S-1   & S-2   &  0 -Unchanged  \\
Mixed 1Rand       & G-1   & G-2   & G-3   &  1 - Merged      \\
Noisy 1Rand       & S-0   & S-0   & S-3   &  2 - Split      \\
Mixed Noisy 1Rand & S-1   & S-2   & S-3   &  3 - Random         \\
2Rand             & G-0   & G-3   & G-3   &       \\
\cmidrule(r){1-4}
\end{tabularx}
\centering
\caption{\textbf{Modularity Problems for Evaluating Similarity Integration Methods}. 2500 samples are split into 10 equally sized clusters, with three modalities are generated for each modality problem. Each modality $X_i$ is characterised by a distribution --- Gaussian (G), Student's-t (S), or Categorical (C) --- and by cluster information: 0) $y_i$ identical to the ground truth $y$, 1) $y_i$ with 5 clusters merged from $y$, 2) $y_i$ produced by splitting $y$ into 20 sub-clusters, and 3) $y_i$ containing 10 random, equally sized clusters unrelated to $y$. These variations allow for a comprehensive assessment of how well similarity integration methods can handle different cluster structures and data distributions.}
\label{tab:3modproblems}
\end{table}

\subsection{Partial Modalities}
We evaluate the effect of partial modality on the similarity integration methods. To include entities with missing modalities, we adapt the methods in the following ways

\begin{itemize}
    \item \textbf{Similarity Network Fusion (SNF)} --- For each pairwise modality distance $S^{(K)}$, the pairwise value between a node $i$ with $NaN$ in $X_k$ and any node $j$ is set to max distance/dissimilarity for that modality. SNF is then computed as normal with max dissimilarity included.  
    \item \textbf{Neighbourhood Based Multi-Omic Clustering (NEMO)} --- NEMO was developed to analyse partial data. The mean relative similarity for any pair of nodes $i$ and $j$ is computed over the modalities where both nodes have recorded data.
    \item \textbf{Concatenated $X_i$} --- Feature mean value imputation in $X_k$ for all nodes with $NaN$ values. Then distance/similarity calculated as normal.
    \item \textbf{Mean $S_i$ imputing \textit{Max}} --- For each pairwise modality distance $S^{(K)}$, the pairwise value between a node $i$ with $NaN$ in $X_k$ and any node $j$ is set to max distance/dissimilarity for that modality. Mean similarity then computed between a pair of nodes $i$ and $j$ across all modalities.
    \item \textbf{Mean $S_i$ ignoring $NaN$} --- The mean similarity for any pair of nodes $i$ and $j$ is computed  over the modalities where both nodes have recorded data.
    \item \textbf{Extreme Mean} --- Thresholding is performed on the pairwise similarity between nodes with recorded values in the modality. The mean similarity for any pair of nodes $i$ and $j$ is computed  over the modalities where both nodes have recorded data. If all values between $i$ and $j$ are $NaN$ after thresholding (including $NaN$ for where $i$ has no recorded data in a modality) then the dissimilarity is set to max.
\end{itemize}

For the partial data evaluation, we select five of the modality problems in Table \ref{tab:3modproblems} for evaluation; \textit{Easy}, \textit{Mixed Normal}, \textit{1Rand}, \textit{Noisy} and \textit{Mixed Noisy 1Rand}. We create five instances of each modality problem and then mask entities from modalities i) at random and ii) based on cluster membership. We mask a maximum of one modality per entity. We compare the AMI of labels predicted by Leiden and SBM clustering to the both the truth cluster membership $y$ and the list of removed modalities per entity --- $y_{NaN}$.

%------------------------------------------------
\FloatBarrier
\section{Results}\label{3sec:results}
\subsection{Integration Networks}

\subsubsection{Clustering Performance}
\begin{sidewaysfigure*}[!htbp] % Single column figure
	\includegraphics[width=\linewidth]{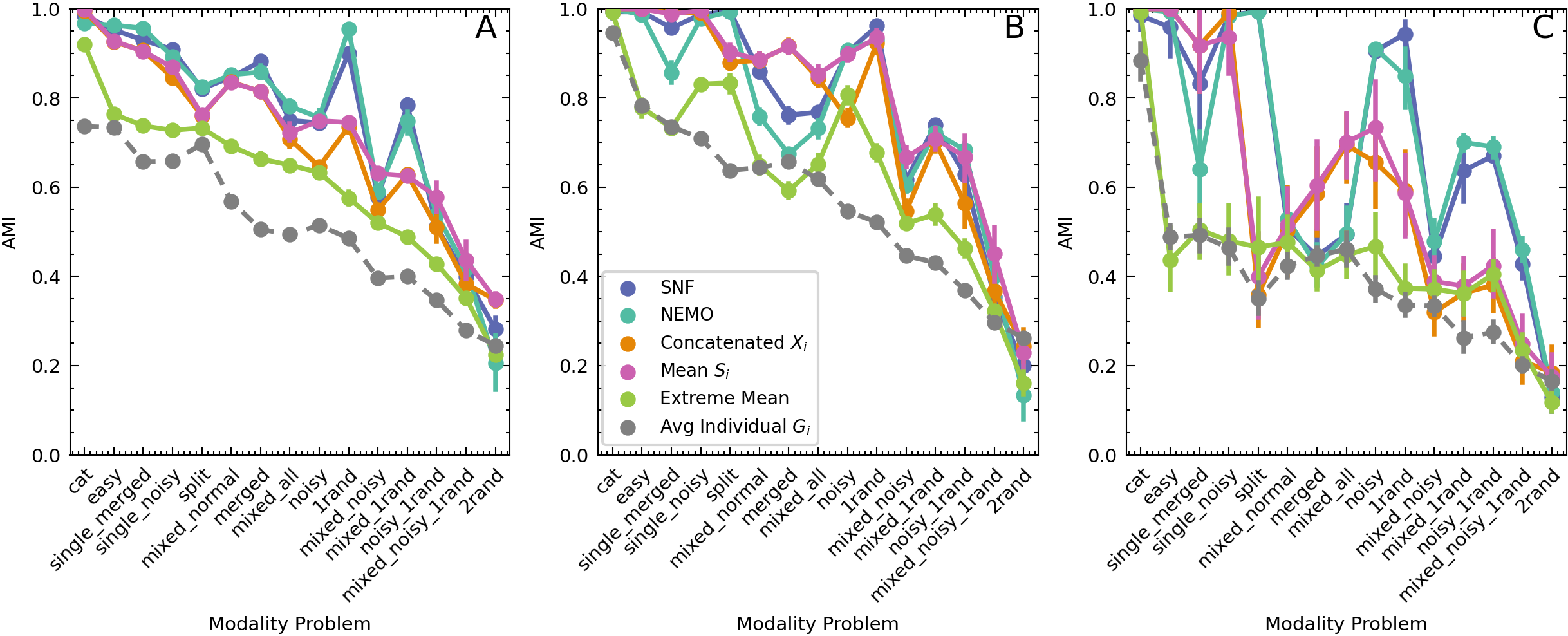}
	\caption{\textbf{AMI Performance Comparison of Similarity Integration Methods Across Multiple Modalities.} AMI performance of A) SBM B) Leiden  and C) Spectral clustering algorithm on 20 instances of 15 different modality problems using Euclidean distance is presented.  Five similarity integration methods are compared: SNF, NEMO, Concatenated $X_i$, Mean $S_i$ and Extreme Mean. The average performance of each clustering algorithm on a KNN network $G_i$ using each individual modality is also shown. We can see all integration methods (including simple concatenation) provide a significant improvement in performance. SNF is consistently outperformed by simpler integration methods such as Mean $S_i$ and NEMO on Leiden clustering. Both NEMO and SNF do offer improvements in the accuracy of SBM and Spectral clustering methods. A network constructed from simple concatenation matches the performance of more complex approaches on easier modality problems. However, in higher noise settings such as \textit{Noisy} and \textit{Mixed Noisy} assessing each modality independently (i.e. using Mean $S_i$, NEMO or SNF) provides an improvement across all clustering algorithms.}
	\label{fig:3modperf}
\end{sidewaysfigure*}

In Figure \ref{fig:3modperf}, the adjusted mutual information (AMI) performance of five similarity integration methods on 20 instances of 15 different modality are shown for A) Stochastic Block Model (SBM), B) Leiden and C) Spectral clustering methods. As a baseline reference, the average performance of the respective clustering algorithm on networks created from each single modality is also shown (\textit{Avg Individual $G_i$}). The modality problems are ordered by the mean performance of all clustering algorithms on individual modality networks.  \\

SBM clustering on SNF and NEMO networks consistently outperforms Mean $S_i$ and Concatenated $X_i$ networks (Figure \ref{fig:3modperf}A). There is very little difference in performance between SNF and Mean $S_i$ for Leiden clustering (Figure \ref{fig:3modperf}B) on more challenging modality problems (Mixed Noisy onwards). SNF performs better on Split clusters and Mean $S_i$ performs better when clusters are merged. There is a significant improvement in the SNF network over the Mean $S_i$ for the Spectral algorithm (Figure \ref{fig:3modperf}C) on more challenging modality problems. 

The most notable differences in clustering performance can be seen on \textit{Split} and \textit{Merged} modality problems. Consider Spectral clustering (Figure \ref{fig:3modperf}C) on \textit{Split}, SNF and NEMO are nearly perfectly accurate where as Concatenated $X_i$ and Mean $S_i$ only match the average performance on individual modality networks. On \textit{Merged}, the opposite can be seen where SNF and NEMO match the average $G_i$ and Concatenated $X_i$ and Mean $S_i$ perform well. The performances on the other clustering algorithms reinforce this behaviour where SNF and NEMO struggle to incorporate merged clusters and Mean $S_i$ and Concatenated $S_i$ struggle with split clusters.

\begin{sidewaystable*}[!htbp]
    \centering
    \begin{tabularx}{\linewidth}{l*{14}{X}}
    \toprule
    Modality Problem & \multicolumn{2}{c}{Easy} & \multicolumn{2}{c}{Single Merged} & \multicolumn{2}{c}{Merged} & \multicolumn{2}{c}{Split} & \multicolumn{2}{c}{1Rand} & \multicolumn{2}{c}{Mixed 1Rand} & \multicolumn{2}{c}{Mixed Noisy} \\
    \cmidrule(lr){2-3} \cmidrule(lr){4-5} \cmidrule(lr){6-7} \cmidrule(lr){8-9} \cmidrule(lr){10-11} \cmidrule(lr){12-13} \cmidrule(lr){14-15}
         & Max & Mean & Max & Mean & Max & Mean & Max & Mean & Max & Mean & Max & Mean & Max & Mean \\
        \midrule
        Graph &  &  &  &  &  &  &  &  &  &  &  &  \\
    SNF & \textbf{0.998} & 0.968 & 0.997 & 0.906 & 0.941 & 0.696 & \textbf{0.998} & \textbf{0.936} & 0.987 & \textbf{0.935} & \textbf{0.837} & 0.720 & 0.674 & 0.546 \\
    NEMO & \textbf{0.999} & \textbf{0.981} & 0.985 & 0.818 & 0.921 & 0.651 & \textbf{0.998} & \textbf{0.937} & 0.972 & 0.911 & 0.794 & \textbf{0.724} & 0.644 & 0.557 \\
    Mean $S_i$ & \textbf{1.000} & 0.976 & \textbf{1.000} & \textbf{0.937} & \textbf{0.983} & \textbf{0.778} & 0.982 & 0.688 & \textbf{0.995} & 0.755 & 0.825 & 0.571 & \textbf{0.797} & \textbf{0.562} \\
    Concatenated $X_i$ & \textbf{1.000} & 0.975 & \textbf{1.000} & \textbf{0.938} & \textbf{0.982} & 0.772 & 0.958 & 0.666 & 0.993 & 0.750 & 0.813 & 0.564 & 0.647 & 0.471 \\
    Extreme Mean & 0.896 & 0.660 & 0.781 & 0.658 & 0.717 & 0.556 & 0.906 & 0.677 & 0.749 & 0.542 & 0.603 & 0.463 & 0.572 & 0.470 \\
    \bottomrule
    \end{tabularx}
    \caption{\textbf{Mean and Maximum AMI Performance Comparison of Similarity Integration Methods Across Multiple Modalities.} Mean and maximum clustering AMI performance on the networks of the five integration methods on 20 instances of several modality problems is shown. We select seven representative modality problems to summarise performance. On problems with multiple merged modalities --- \textit{Single Merged}, \textit{Merged}, \textit{Mixed Noisy}, Mean $S_i$ outperforms SNF and NEMO both in Max and Mean AMI. On \textit{Split}, \textit{1Rand} and \textit{Mixed 1Rand}, SNF, Mean $S_i$'s max performance is quite strong. It is close in performance to SNF on all 3, outperforming it on \textit{1Rand}. Yet its mean clustering performance is significantly worse. The drop in performance is more significant than SNF's corresponding drop on merged clusters.}
    \label{3tab:3modperf}
\end{sidewaystable*}

Extreme Mean has the worst performance on all clustering algorithms. Concatenated $X_i$ is very similar in performance to Mean $S_i$. Perhaps the simplest integration method, Concatenated $X_i$, does not see any drop in performance over the more complex methods on less challenging clustering problems. It is notable that the Concatenated $X_i$ network produced in datasets containing Mixed Student's-t distributed data are significantly worse for clustering. All 3 clustering algorithms show a significant drop in performance compared to Mean $S_i$.

While a simpler method in comparison to SNF, NEMO has very similar performance. Notably, NEMO networks also see a significant drop in performance of the Leiden and Spectral algorithms on Merged clusters. Like SNF, both SBM and Spectral clustering see significant improvement on NEMO networks over Mean $S_i$ and Concatenated $X_i$. NEMO does not handle merged networks as well as SNF and the drop in performance is more significant. \\

In Table \ref{3tab:3modperf}, we summarise the performance of the integration methods by on a subset of the modality problems. The maximum and average AMI performance of the clustering algorithms on the networks of the five similarity integration methods on 20 instances of each problem is shown. Reinforcing Figure \ref{fig:3modperf}, the maximum performance of SNF and Mean $S_i$ is closely matched across all problems. There is far greater variation in the average performance --- on \textit{Split}, \textit{1Rand} and \textit{Mixed 1Rand} the average performance of Mean $S_i$ is significantly lower. \textit{Single Merged}, \textit{Merged} and \textit{Mixed Noisy} all contain multiple merged cluster modalities and we can see reduction in SNF and NEMO performance is consistent both in maximum and average performance. 

\subsubsection{Network Properties}
\begin{figure*}[!htbp] % Single column figure
	\includegraphics[width=0.8\linewidth]{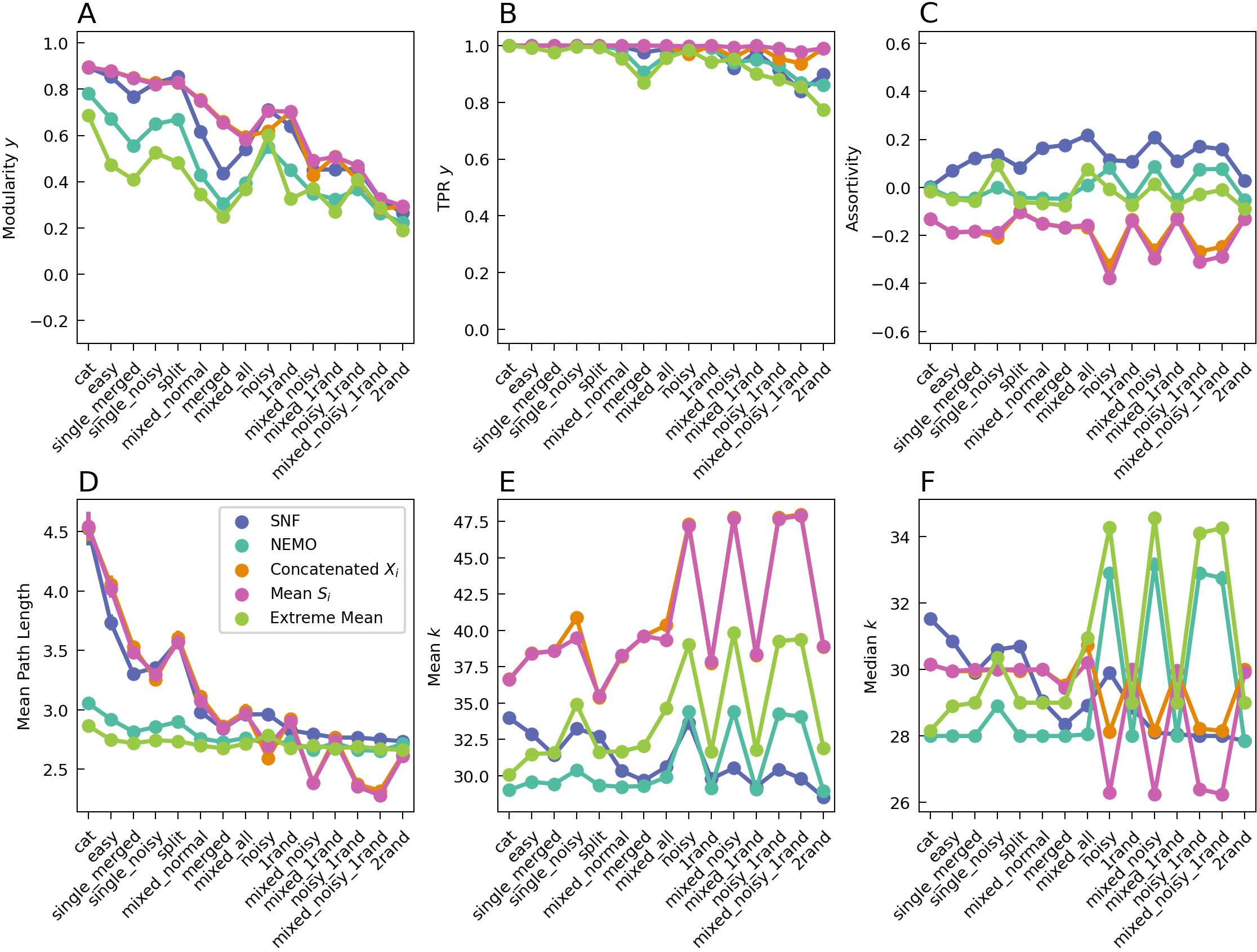}
    \centering
	\caption{\textbf{Comparison of the Network Properties of Integration Methods.} The A) Modularity $y$, B) TPR $y$, C) Assortativity, D) Mean path length, E) Mean Degree and F) Median Degree are shown for 20 instances of networks on all 15 modality problems. Mean $S_i$ and Concatenated $X_i$ have very similar properties, with Mean $S_i$ slightly more modular (A) and more likely to contain edges between high and low degree nodes (C). Unlike other methods, SNF structure is less affected by Mixed Student's-t distributed data (D-F). Its density does not increase and the mean path length is consistent. From C), we can see SNF has positive assortativity --- connections are more likely between nodes of similar degree. NEMO networks are neutral and connections between nodes of all degrees are equally likely.}
	\label{fig:3modgraphcomp}
\end{figure*}

In Figure \ref{fig:3modgraphcomp}, several properties of the networks produced by the similarity integration methods are shown for the 15 modality problems. The change in A) Modularity of true clusters $y$, B) Triad Participation Ratio (TPR) of true clusters $y$, C) Assortativity, D) Mean Path Length, E) Mean Degree and F) Median degree across the modality problems are shown. For all integration methods, a KNN network with $K=25$ and 2500 nodes is constructed.  

We can see Mean $S_i$ and Concatenated $X_i$ produces the most modular networks (\ref{fig:3modgraphcomp}A). They are significantly more modular than SNF networks on problems containing multiple merged modalities; Single Merged, Merged, Mixed Normal and Mixed All. Extreme Mean shows an increase in Modularity relative to other network on problems containing Mixed Student's-t data; Single Noisy, Noisy, Mixed Noisy and Noisy 1Rand. SNF networks are as modular as the Mean $S_i$ and Concatenated $X_i$ networks on data without merged clusters. NEMO is significantly less modular than these networks on all modality problems. 

In Figure \ref{fig:3modgraphcomp}B the TPR rate is consistently high for all methods with the exception of Extreme Mean. This indicates there is strong internal connectivity within the clusters. Nearly all nodes are triads and it is only on the more challenging noisy modality problems where the rate of triads within clusters begins to drop. As the modality problems increase in difficulty, all methods show a decrease in TPR. Mean $S_i$ is the most resistant and is consistently high.

From Figure \ref{fig:3modgraphcomp}C, we can see SNF networks have positive degree assortativity coefficients on nearly all modality problems. The correlation is not extremely strong but on average connections between nodes of the same degree are more likely than connections between high and low degree nodes. In contrast,  Mean $S_i$ and Concatenated $X_i$ networks have negative assortativity but the strength of the correlation is not very strong. NEMO and Extreme Mean show neutral assortativity for most modality problems and within these networks connections between all types of node degree are equally likely. One notable pattern is the drop in assortivity shown by Mean $S_i$ and Concatenated $X_i$ networks on modality problems containing noisy modalities. In contrast, NEMO networks show an increase in degree assortativity on these problems.

Figure \ref{fig:3modgraphcomp}D shows the Mean Path Length for all networks drops as the modality problems becomes more challenging. NEMO and Extreme are the most consistent but this are result of the high interconnectivity i.e. lower mean path length on easier problems. An decrease in mean path length corresponds to clusters becoming less distinct as more inter cluster edges are present in the network. The more connections between clusters the lower the average path length as the network becomes easier to traverse. Mean $S_i$ and Concatenated $X_i$ show a significant drop in mean length on data containing Mixed Student's-t distributed modalities. SNF, NEMO and Extreme Mean are more resistant to the noisy data and do not show a decrease. 

The same number of neighbours ($K=25$) are assigned to each node in each network. As a result, any increase in mean degree i.e. an increase in network density and total number of edges in the network, implies that less nodes are mutual nearest neighbours. Mutual nearest neighbours are nodes which include each other in their set of nearest neighbours (NN).  A drop in density occurs when nodes are mutual NN because only one single edge is added to the network instead of the two edges that would exist if they were not mutual NNs. As seen in Figure \ref{fig:3modgraphcomp}E, All networks except SNF show an increase in mean degree on modality problems containing Mixed Student's-t data. Mean $S_i$ and Concatenated $X_i$ consistently have the highest density of all networks.

Very different behaviours occur in the median degree of the distribution however. From Figure \ref{fig:3modgraphcomp}F we can see Mean $S_i$ and Concatenated $X_i$ display a decrease in median degree where NEMO and Extreme Mean show an increase. This can be explained by the change in assortativity on the networks. In NEMO and Extreme Mean the additional edges that result in an increase in density occur between nodes of similar degree. In Mean $S_i$ and Concatenated $X_i$, these connections are between high and low degree nodes. When we consider the corresponding decrease in mean path length seen on these networks, these edges likely occur between clusters rather than within clusters. \\

\FloatBarrier
\subsection{Effect of Partial Data}\label{3sec:results:partial}

\subsubsection{Clustering Performance}
\begin{figure*}[!htbp] % Single column figure
	\includegraphics[width=0.8\linewidth]{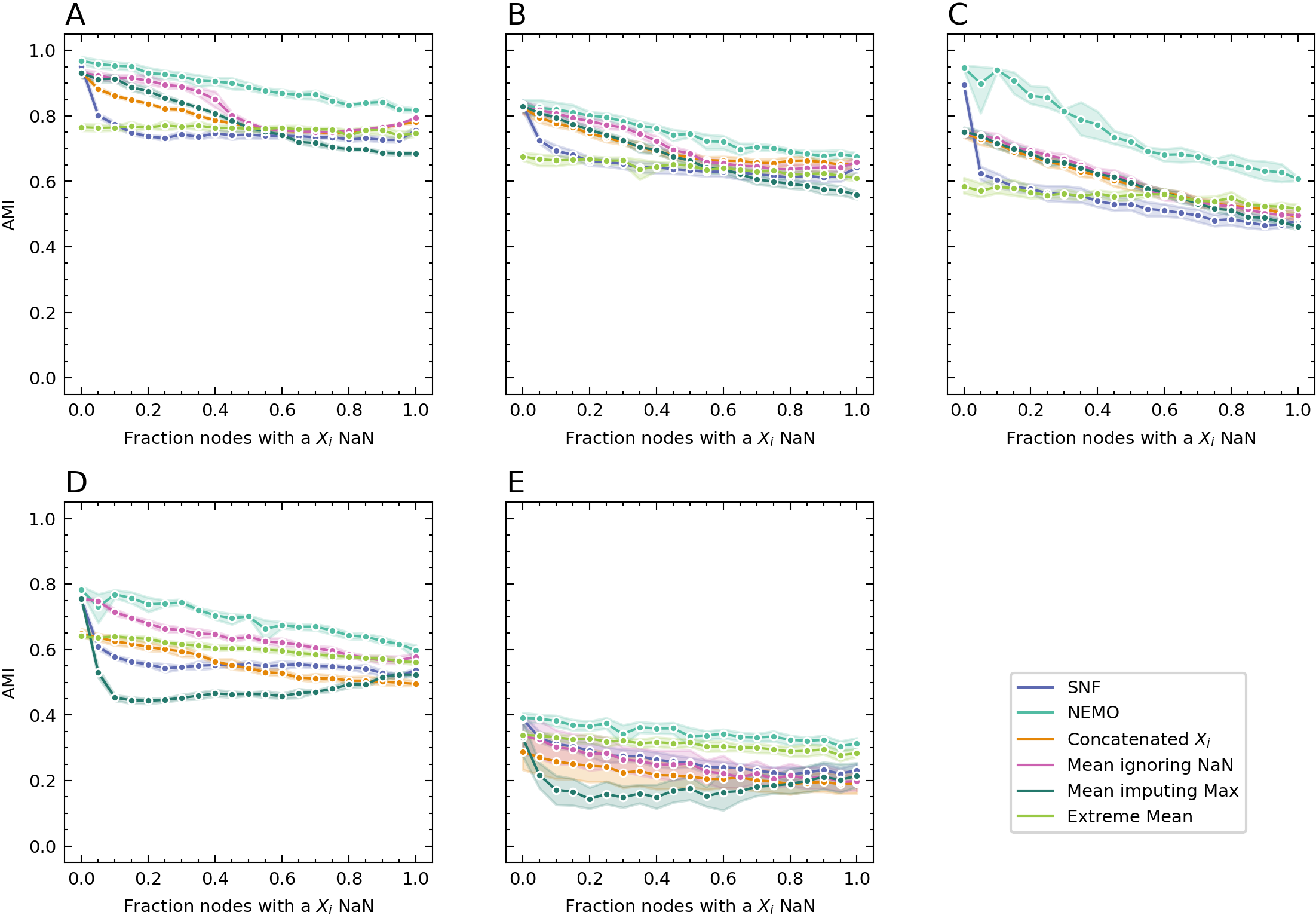}
    \centering
	\caption{\textbf{Comparison of AMI Performance of Integration Methods on Data Partial At Random.} Change in SBM AMI performance for data partial at random on 5 instances of A) \textit{Easy}, B) \textit{Mixed Normal}, C) \textit{1Rand}, D) \textit{Noisy} and E) \textit{Mixed Noisy 1Rand} modality problems. Extreme Mean is the least affected by partial data across all modality problems showing little to no change in performance. Mean ignoring $NaN$ is more resistant to partial data than other methods up to a certain level of partial data before dropping in performance (A and B). SNF is highly sensitive to partial data and initially shows a significant drop in performance but is stable thereafter. Mean imputing Max performance degrades quickly with partial data in Noisy modality problems (D and E).}
	\label{3fig:pmod_random_perf}
\end{figure*}

Figure \ref{3fig:pmod_random_perf} illustrates the variation in SBM AMI performance for increasing fraction of nodes with data partial at random across five instances of A) \textit{Easy}, B) \textit{Mixed Normal}, C) \textit{1Rand}, D) \textit{Noisy}, and E) \textit{Mixed Noisy 1Rand} modality problems.  NEMO emerges as the most resilient method to partial data. It displays the lowest reduction in performance across all five modality problems. The performance of SNF degrades significantly with any inclusion of partial data. Surprisingly, as the level of partial data increases the performance does not degrade further. Mean ignoring $NaN$ initially is resistant to data partial at random in the Easy and Mixed Normal modality problems (Figure \ref{3fig:pmod_random_perf}A \& B) but once a threshold of partial data is crossed its performance drops. On the more challenging modality problem it displays a consistent reduction for all levels of partial data (Figure \ref{3fig:pmod_random_perf}C-E). 

\begin{figure*}[!htbp] % Single column figure
	\includegraphics[width=0.8\linewidth]{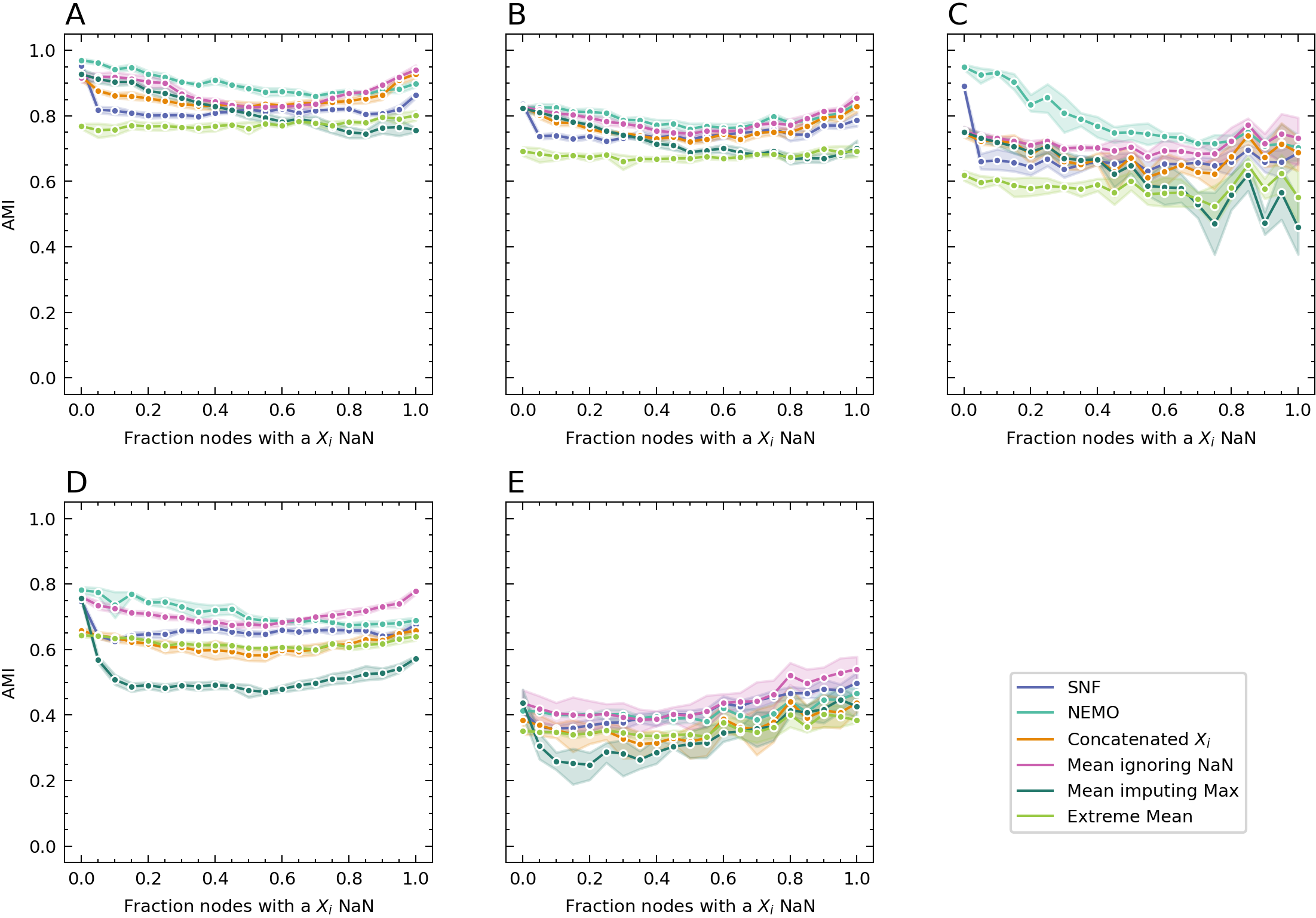}
    \centering
	\caption{\textbf{Comparison of AMI Performance of Integration Methods on Data Partial Based on Cluster.} Change in SBM AMI performance for data partial based on cluster on 5 instances of A) \textit{Easy}, B) \textit{Mixed Normal}, C) \textit{1Rand}, D) \textit{Noisy} and E) \textit{Mixed Noisy 1Rand} modality problems. As the fraction of nodes with partial data increases, the clusters in each modality become more consistent. The effect of partial data is most severe at 50\% when the enough members of the cluster remain to add noise to the pairwise similarity within a modality but not enough to form a strong cluster. When 100\% of nodes have a $NaN$ $X_i$, we only have two measurements of pairwise similarity from the modalities. This explains the increased noise of all methods in $1Rand$ (C) at higher levels of partial data --- for a majority of nodes half of the similarity measurements are completely random. }
	\label{3fig:pmod_clusterb_perf}
\end{figure*}

Figure \ref{3fig:pmod_clusterb_perf} shows the change in SBM clustering AMI for cluster based partial data across five instances of modality problems: A) \textit{Easy}, B) \textit{Mixed Normal}, C) \textit{1Rand}, D) \textit{Noisy}, and E) \textit{Mixed Noisy 1Rand}. The performance of methods demonstrates improvement on certain modalities with cluster based partial data. As the fraction of nodes with partial data increases, the consistency of clusters within each modality improves. The impact of partial data is most pronounced at 50\% when there are enough members in the cluster to introduce noise to pairwise similarity within a modality, but not sufficient to form a robust cluster. When 100\% of nodes have a $NaN$ $X_i$, it results in only two measurements of pairwise similarity from the modalities. This explains the heightened noise observed in all methods in the \textit{1Rand} modality (Figure \ref{3fig:pmod_clusterb_perf}C) at higher levels of partial data. For the majority of nodes, half of the similarity measurements are entirely random under these conditions.

In this setting, all methods exhibit increased resilience to partial data compared to data partial at random. The improvement of NEMO over other methods is significantly reduced. Mean ignoring $NaN$ particularly demonstrates highly improved performance. Although there is notably more variance in the performance of methods with cluster based partial data over data partial at random, this variance increase aligns with consistently higher performance.

Mean ignoring $NaN$ consistently shows increased performance for 100\% partial data compared to no partial data. Other methods also display similar performance increases though no to the same extent. At 100\% partial data, entire clusters have been removed from different $X_i$, resulting in each modality containing fewer clusters. In our underlying data generation procedure we strategically place clusters close enough to one another to ensure overlap. Consequently, when clusters are removed, the increased distance between clusters facilitates easier distinction. Additionally, in merged data, clusters are no longer combined together, making the remaining clusters easier to identify.

\subsubsection{Relationship to missing labels}

\begin{figure}[!htbp] % Single column figure
	\includegraphics[width=\linewidth]{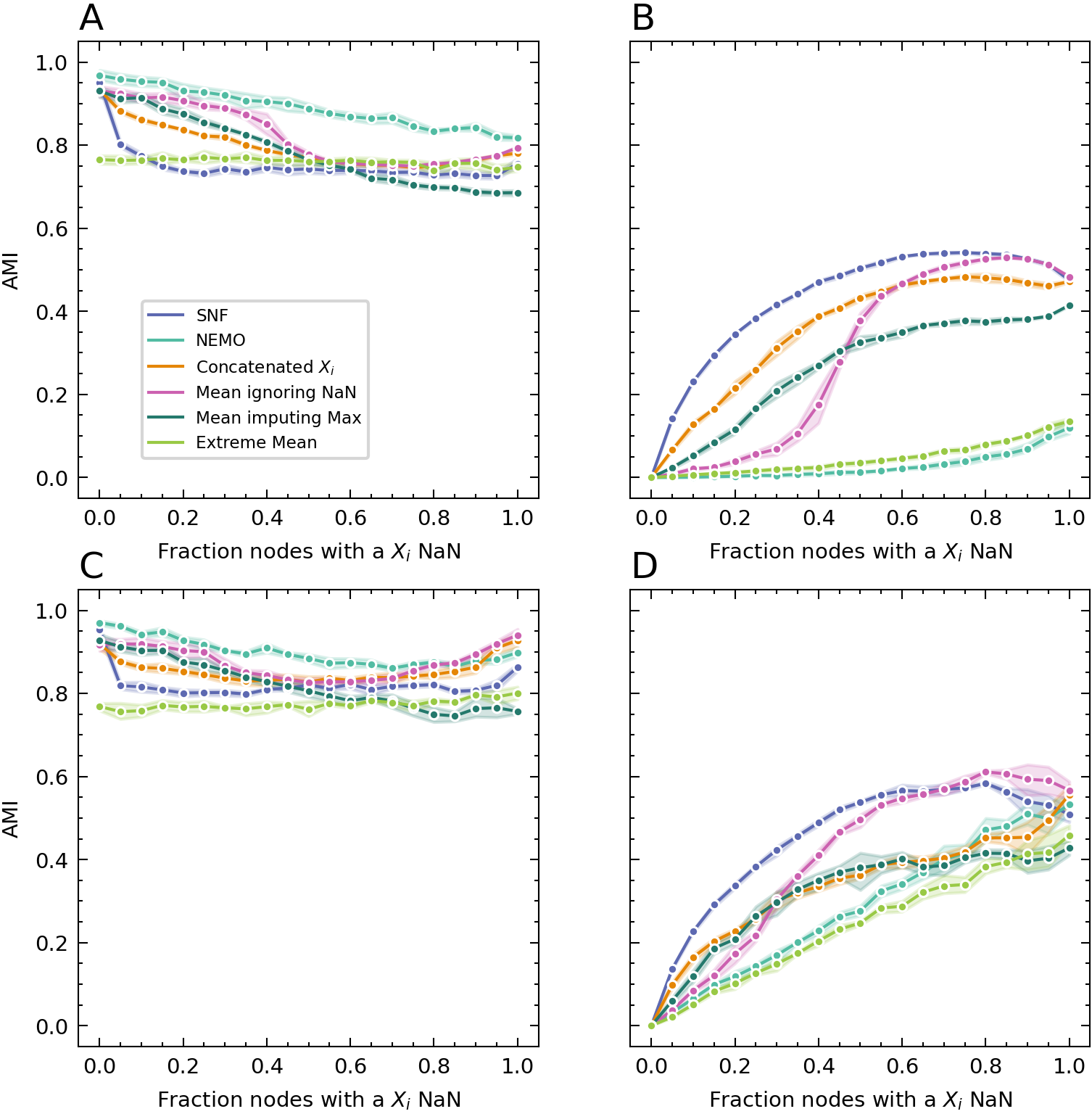}
    \centering
	\caption{\textbf{ SBM AMI Between $y$ and $y_{NaN}$ on the \textit{Easy} Modality Problem With Increasing Partial Data.} SBM AMI between $y$ and $y_{NaN}$ on five instances of \textit{Easy} modality problem with data partial at random and based on cluster. We show A) $y$ partial at random, B) $y_{NaN}$ partial at random, C) $y$ partial based on cluster and D) $y_{NaN}$ partial based on cluster. SNF is the most significantly affected by partial both at random and based on cluster. Mean ignoring $NaN$ experiences a change in resistance when around 50\% of individuals are absent at random from an $X_i$. It rapidly drops in performance and becomes more similar to $y_{NaN}$. Concatenated $X_i$ and Mean imputing Max quickly deteriorate in performance and similar to SNF quickly align with $y_{NaN}$}
	\label{3fig:pmod_ynan_y}
\end{figure}

Figure \ref{3fig:pmod_ynan_y} shows the SBM AMI between $y$ and  $y_{NaN}$ on five instances of \textit{Easy} modality problem with data missing at random and based on cluster. We show A) $y$ partial at random, B) $y_{NaN}$ partial at random, C) $y$ partial based on cluster and D) $y_{NaN}$ partial based on cluster. $y_{NaN}$ AMI measures the agreement between the list of modalities each individual is absent from and the discovered clusters. The higher this AMI is the more influence partial data has on the clustering process. We can see SNF's initial drop in $y$ AMI performance corresponds to a significant increase in $y_{NaN}$ AMI for data both partial at random and cluster based. While the similarity between $y_{NaN}$ and the predicted clusters increases, the $y$ AMI remains consistent. The transition in $y$ clustering performance of Mean ignoring $NaN$ on data partial at random (Figure \ref{3fig:pmod_ynan_y}A) is amplified in $y_{NaN}$ and the corruption of the cluster structure due to partial data is clearly visible (Figure \ref{3fig:pmod_ynan_y}B). Surprisingly, Mean imputing Max's $y_{NaN}$ AMI is lower than Mean ignoring $NaN$ at higher levels of partial data despite the worse $y$ performance. This is true for both data partial at random and cluster based (Figure \ref{3fig:pmod_ynan_y}B \& D). 

NEMO and Extreme Mean show very interesting behaviour in their $y_{NaN}$ AMI. Both are highly resistant to data partial at random and bear little resemblance to the list of absent modalities of each individual (Figure \ref{3fig:pmod_ynan_y}B). Yet for cluster based partial data (Figure \ref{3fig:pmod_ynan_y}D), they display a steady increase in $y_{NaN}$ AMI. NEMO and Extreme Mean offer potential measures for detecting whether partial data is related to underlying clusters within the data --- a low resemblance between clusters detected on NEMO and the labels of absent modalities could be indicative data is partial at random. Further investigation is required but the significant difference in behaviour of these methods across the types of partial data is promising.

% \FloatBarrier
\section{Discussion}\label{3sec:discussion}

% implications of chapter
SNF does not emerge from this analysis as the clear choice of integration method. On key modality problems involving merged clusters, it is outperformed by simpler approaches such as Mean $S_i$ (Figure \ref{fig:3modperf} \& Table \ref{3tab:3modperf}). Merged clusters are particularly pertinent for disease subtype analysis where our expectation is that there will be a significant amount of overlap between subtypes on several modalities. For the integration of partial modalities, again a simpler approach, NEMO, is a far more optimal choice of algorithm (Figure \ref{3fig:pmod_random_perf} \& \ref{3fig:pmod_clusterb_perf}).  The diffusion approach of SNF does show improvements in a number of scenarios. Most notably, the incorporation of random modalities (1Rand) and Split clusters. Furthermore, its mean clustering performance is more consistent with SBM and Spectral clustering performing better across most modality problems on SNF networks than Mean $S_i$ (Figure \ref{fig:3modperf} \& Table \ref{3tab:3modperf}).

Consistency of pairwise similarity scores across modalities emerges as the single most important factor differentiating integration method performance. The benefit and drawbacks of KNN selection in SNF and NEMO is highlighted by differences in consistency. Mean $S_i$ is more optimal in \textit{Merged} clusters. In \textit{Merged} clusters, two ground truth clusters are combined and the points of both clusters are placed at random around a single cluster centre. On average the similarity of members of a cluster is quite high. But while the average similarity is high, the $K$ nearest neighbours of a particular node are just as likely to contain nodes from another cluster as its own. Across multiple modalities the increased similarity to nodes in the second cluster will drop while the similarity to nodes within a cluster will remain high. Mean $S_i$ successfully identifies this. SNF and NEMO however do not calculate similarity using nodes that are not in a node's $K$ nearest neighbours. With multiple merged clusters, the neighbours of a node selected in each modality will consistently contain nodes from other clusters increasing the difficulty of successfully identifying nodes within a cluster.

On the other hand, SNF and NEMO show strong performance on \textit{Split} modality problems. In \textit{Split}, clusters are separated apart. A node's $K$ nearest neighbours will contain members of its cluster. Across different modalities, the particular neighbours might change but all will originate from the same cluster. For Mean $S_i$, the similarity between nodes within a sub-cluster will remain high but for the rest of the sub-cluster there will be low similarity. As we aggregate across modalities, the similarity between nodes within a cluster will oscillate between high and low reducing their overall similarity and increasing the difficulty of connecting nodes within a cluster. 

It must be noted that the addition of corruptive modalities containing unrelated community structure (random cluster information) can have a significant impact on the ability of integration methods to detect communities robustly. If we consider the original modality problems using three modalities --- \textit{Easy}, \textit{1Rand} and \textit{2Rand} (Figure \ref{fig:3modperf}). There is a slight drop in performance of most methods on \textit{1Rand} but it is not significant. The random modality is successfully incorporated. However, on \textit{2Rand} the performance drops significantly. While the \textit{2Rand} example is extreme, it is illustrative of the dangers of including additional modalities. Not all modalities will necessarily contain the community information we are seeking to identify.

SNF struggles significantly with partial data. It is highly sensitive to partial data and shows a significant drop in performance with its inclusion (Figures \ref{3fig:pmod_random_perf} \& \ref{3fig:pmod_clusterb_perf}). In contrast, NEMO is a method developed with partial data in mind and the benefit shows. It is highly resistant to partial modalities and shows the lowest drop in performance as the rate of partial data increases. Within the methods shown here, partial data strategies that focus only on shared modalities and avoid punishing increased uncertainty show more success --- Mean Ignoring $NaN$, NEMO and Extreme Mean. Interestingly Mean Ignoring $NaN$ begins to struggle after a threshold of partial data is reached. The methods that filter similarity values prior to aggregating, Extreme Mean and NEMO, are more consistent across all levels of partial data.

The differences in behaviour between cluster based partial data and data partial at random is highly remarkable. Most notably certain methods improve in performance with cluster based partial data over their complete versions (Figures \ref{3fig:pmod_random_perf} \& \ref{3fig:pmod_clusterb_perf}). This has significant implications. As discussed in Section \ref{3sec:intro}, partial data is typically removed from analysis. Yet here are a number of scenarios where increased partial data is highly beneficial. We should caution here that there is likely a simplification of the clustering problem that occurs with partial that is a result of our synthetic data generation process (less clusters in a modality results in higher separation between the remaining clusters). A more in depth examination comparing the clustering with partial data to clustering on the set of data with complete measurements across modalities is needed. In any case, the reasons for partial data can be complex and, more importantly, the optimal methods for incorporating partial data can change based on the partial data process. For example, Mean $S_i$ outperforms NEMO at high levels of cluster based partial data as seen in Figure \ref{3fig:pmod_clusterb_perf}.

\subsection{Limitations}
It must be acknowledged that our use of synthetic data introduces limitations. Our synthetic data generation process is constrained by our assumptions and lacks realistic complexity. These factors may potentially compromise the generalisability of our findings. The data distributions used to embed clusters are relatively simple and do not contain complex interactions between clusters. While the addition of high noise distributions like mixture of Student's-t introduces outliers and increased difficulty, there is no guarantee this is reflective of real world challenges. 

There is limited variety in how clusters are embedded across modalities. The variation in cluster structures is highly simplified. While some implications can be gleaned from our merged and split clusters, these simplifications do not encompass the full spectrum of possibilities in real-world scenarios. Understanding the sensitivity of integration methods to the specific combination of modalities used is crucial for practical applicability.

Comparisons with multi-omic data sources reveal further limitations. Our synthetic data maintains consistent cluster distributions with limited variations in consistency across modalities. There are no changes in the number of features, and the data remains relatively low-dimensional compared to real multi-modal data, where modalities may have tens of thousands to hundreds of thousands of features.

There are additional limitations in our partial data analysis. Our choice of imputation is highly conservative, prioritising the penalisation of individuals with missing data to reflect the increased uncertainty in their features. This approach likely explains sensitivity of Similarity Network Fusion (SNF) to partial data. A less punishing imputation strategy might enhance SNF's performance. Additionally, our exploration of partial data is quite constrained, as each individual is at most missing from one modality. Real world datasets individuals are missing from several modalities. We also do not examine the effect of partial data on local structure or determine when cluster information collapses.

\subsection{Future Work}
In terms of future research, a key direction would involve expanding the modality generation framework. The consistency of a pair of nodes' pairwise similarity and similarity to their wider neighbours within a cluster are essential factors determining the addition of edges between nodes in the network. With this in mind exploring diverse configurations for cluster information across modalities is crucial. For instance, introducing targeted random noise or increasing pairwise swaps of features could be a method to test new types of consistency. Another avenue could involve adding fully random modalities without embedded clusters, diverging from the current approach of introducing random clusters. The primary goal is to systematically delve into pairwise similarity consistency in a targeted manner, enhancing our understanding of when specific methods prove more effective.

An additional enhancement to our analysis would be a concerted effort to better mirror real-world multi-omic and multi-modal data. Introducing variations in the number of features, cluster size, and cluster distribution would increase the realism of our synthetic data, making it more representative of the intricacies found in real-world datasets and significantly enhancing the generalisability of our framework.

Another avenue of investigation is the assessment of partial data using less conservative imputation strategies, and numerous sophisticated methods exist for imputing \textit{item non-response}. If we consider our similarity measurements for each modality as a set of features, we can leverage more complex imputation strategies, potentially leading to improved performance. This is particularly relevant for methods like Similarity Network Fusion (SNF), which exhibited high sensitivity to partial data, making them potential beneficiaries of such strategies.

Moreover, a deeper investigation into the effects of partially complete modalities would be highly beneficial. Through the application of more complex partial modality strategies, we can better mirror the types of incompleteness observed in real-world data. Our strategy for introducing partial modalities factors that relate to the underlying clusters is relatively simple --- a realistic factor to likely to be more complex. Furthermore, in real world datasets, individuals may be absent from several modalities. Is there a threshold where the level of absent modality data renders an individual more of a hindrance than a benefit? To what extent do partial individuals corrupt their neighbours? Do some methods handle increased partial data more effectively than others?

Lastly, a valuable future direction involves comparing network clustering approaches to other multi-modal clustering methods, such as dimensionality reduction and matrix factorisation methods. Beyond just a comparison of clustering accuracy, exploring whether networks created from the embeddings produced by these alternative methods are more informative and reflective of community structures would provide valuable insights.

\FloatBarrier
%----------------------------------------------------------------------------------------
%	 REFERENCES
%----------------------------------------------------------------------------------------

\printbibliography % Output the bibliography

\end{document}